\documentclass[aps,prl,reprint, nofootinbib]{revtex4-1}
\usepackage{amsmath}
\usepackage{amssymb}
\usepackage{amsthm}
\usepackage{MnSymbol}
\usepackage{nicefrac}
\usepackage{amsfonts}
\usepackage{dsfont}
\usepackage[hidelinks]{hyperref}  

\usepackage[markup=nocolor]{changes}

\usepackage{graphicx}

\usepackage[normalem]{ulem}

\usepackage{dcolumn}
\usepackage{bm}

\usepackage{color}

\newcommand{\bea}{\begin{eqnarray}}
\newcommand{\eea}{\end{eqnarray}}
\newcommand{\be}{\begin{equation}}
\newcommand{\ee}{\end{equation}}

  \graphicspath{{../Figures/}}

\newcommand{\Int}{\int\!\!\!\mathrm{d}^4x\sqrt{g}\,} 
\newcommand{\Intb}{\int\!\!\!\mathrm{d}^4x\sqrt{\bar{g}}\,} 

\begin{document}

\title{Application of positivity bounds in asymptotically safe gravity}

\author{Astrid Eichhorn}
\email{eichhorn@sdu.dk}
\affiliation{CP3-Origins, University of Southern Denmark, Campusvej 55, DK-5230 Odense M, Denmark} 
\author{Marc Schiffer}
\email{mschiffer@perimeterinstitute.ca}
\affiliation{Perimeter Institute for Theoretical Physics, 31 Caroline Street North, Waterloo, ON N2L 2Y5}
\author{Andreas Odgaard Pedersen}
\email{anpe519@student.sdu.dk}
\affiliation{CP3-Origins, University of Southern Denmark, Campusvej 55, DK-5230 Odense M, Denmark}

\begin{abstract}
Positivity bounds are bounds on the Wilson coefficients of an effective field theory. They hold, if the ultraviolet completion satisfies unitarity, microcausality, locality and Lorentz symmetry; accordingly their violation signals a violation of at least one of these properties of the ultraviolet completion. We explore whether positivity bounds on four-photon-couplings hold, when the ultraviolet completion is an asymptotically safe gravity-photon theory. By working at  sixth order in an expansion in the electromagnetic field strength,
we discover indications that positivity bounds hold for effective field theories
 that are UV completed by the  asymptotically safe Reuter fixed point. We also perform various tests of the robustness of our result.
 This amounts to a nontrivial and critical indication of the physical viability of asymptotically safe gravity. 
 \end{abstract}

\maketitle

\emph{Motivation:} 
In this work, we apply positivity bounds \cite{Adams:2006sv,Pham:1985cr,Pennington:1994kc} in asymptotically safe quantum gravity as a litmus test of causality and unitarity in the theory.
Asymptotically safe quantum gravity \cite{Weinberg:1980gg,Reuter:1996cp} is a quantum field theory of gravity. It is based on an enhanced symmetry at small, sub-Planckian distance scales, namely quantum scale symmetry.  This symmetry is realized at a fixed point of the Renormalization Group (RG) flow.
Such a fixed point addressed the 
problem of perturbative non-renormalizability, which consists of a loss of predictivity through introducing infinitely many counterterms. A fixed point generates relations between the corresponding, infinitely many couplings of the theory. These relations persist down to the infrared (IR), if we start from a quantum scale-symmetric theory in the ultraviolet (UV). This renders the theory predictive despite the presence of higher-order interactions which correspond to the counterterms of perturbative renormalization. 
For gravity, such a fixed point is interacting and known as the \emph{Reuter fixed point}. 
Following a proposal by Weinberg \cite{Weinberg:1980gg} and early evidence in 2+$\epsilon$ dimensions \cite{Gastmans:1977ad, Christensen:1978sc, Kawai:1989yh}, see also \cite{Martini:2021slj, Martini:2022sll, Martini:2023qkp} for recent studies, 
a breakthrough came with Reuter's adaptation of functional Renormalization Group techniques \cite{Wetterich:1992yh, Morris:1993qb, Ellwanger:1993mw} to quantum gravity \cite{Reuter:1996cp}. With these techniques, compelling evidence for the requisite interacting RG fixed point has been collected, both in pure gravity and in gravity-matter theories, see \cite{Souma:1999at,Lauscher:2001ya,Reuter:2001ag,Lauscher:2002sq,Litim:2003vp,Codello:2006in,Machado:2007ea,Codello:2008vh,Benedetti:2009rx} for early evidence and \cite{Falls:2020qhj, Baldazzi:2021orb, Knorr:2021slg, Kluth:2022vnq, Baldazzi:2023pep, Saueressig:2023tfy, Korver:2024sam,Burger:2019upn, Pastor-Gutierrez:2022nki}
and references therein for the most recent evidence, as well as  \cite{Eichhorn:2018yfc,Bonanno:2020bil,Eichhorn:2022gku,Saueressig:2023irs, Pawlowski:2023gym, Knorr:2022dsx, Morris:2022btf,Wetterich:2022ncl, Platania:2023srt} for recent reviews.\\
There is a critical open question, namely whether the theory is physically viable in that it respects causality and unitarity. Answering this question is technically non-trivial due to
the effect of higher-derivative interactions. These are unavoidably present at an interacting fixed point, see, e.g., \cite{Lauscher:2002sq,Codello:2008vh,Benedetti:2009rx,Falls:2013bv,Gies:2016con,Denz:2016qks,Falls:2017lst,Falls:2020qhj,Baldazzi:2023pep}, but may endanger the status of the theory as a viable physical theory. 
They may introduce 
ghost degrees of freedom. Ghosts are generically present in simple examples with non-degenerate Hamiltonians and finite many higher-order derivatives, which result in Ostrogradsky instabilities \cite{Woodard:2015zca}. 
Such instabilities can be traded for unitarity-violation in the quantum theory. Going beyond the simplest examples, the situation becomes more subtle and examples of healthy theories with higher-order time derivatives and stable time evolution are known, see, e.g., \cite{Barnaby:2007ve,Gleyzes:2014dya,Langlois:2015cwa,Buoninfante:2020ctr} for different classes of such theories. There are even examples in which ghost degrees of freedom do not result in runaway instabilities in the time evolution of the classical theory \cite{Deffayet:2021nnt,Deffayet:2023wdg}. At the quantum level, it has been shown how ghosts can be avoided and, e.g., traded for potentially harmless violations of microcausality \cite{Donoghue:2019fcb, Donoghue:2019ecz,Donoghue:2021eto}. This mechanism may be relevant for quadratic gravity, see also \cite{Anselmi:2017lia, Anselmi:2018kgz, Anselmi:2018tmf} for a related proposal and \cite{Held:2021pht,Held:2023aap} for results on the well-posedness of the equations of motion at the classical level despite fourth-order time derivatives.
Against this background, it would be premature to dismiss asymptotically safe gravity as a physically viable theory based on the presence of higher-derivative terms. In fact, it has been pointed out that the analytic structure of the propagator might not have ghost poles 
\cite{Bosma:2019aiu,Knorr:2021niv, Bonanno:2021squ,Fehre:2021eob} and scattering amplitudes might obey unitarity bounds \cite{Draper:2020bop}.
Studies that employ truncations of higher-derivative terms to finite order can, however, induce spurious ghosts in the propagator even in theories which are known to be unitarity \cite{Platania:2020knd, Platania:2022gtt}. This makes the assessment of unitarity challenging when working to finite order in a truncation of the gravitational dynamics.

In this work, we pioneer\footnote{During the course of this project, we became aware of the related work \cite{ABtoAppear} by B.~Knorr and A.~Platania, whom we thank for discussions.} a different way of probing the physical viability of the theory: we use positivity bounds to place new constraints on asymptotically safe gravity.

\emph{Positivity bounds:} Positivity bounds are bounds on the Wilson coefficients in an effective field theory. They are derived by demanding unitarity, locality, microcausality, and Lorentz invariance, but 
otherwise remaining agnostic
about the ultraviolet (UV) completion \cite{Adams:2006sv,Pham:1985cr,Pennington:1994kc}. Wilson coefficients parameterize the scattering amplitudes in the theory, and unitarity, locality, microcausality and Lorentz invariance in turn constrain the dependence of the scattering amplitudes on the Mandelstamm variables. For instance, Lorentz invariance implies 
crossing symmetry of the amplitude; unitarity and causality restrict the analytic structure of the amplitude. Positivity bounds arise as part of a broader S-matrix bootstrap, which aims at mapping the space of S-matrices of all consistent quantum field theories \cite{Paulos:2016fap,Kruczenski:2022lot}, see also \cite{EliasMiro:2022xaa} for the relation of S-matrix bootstraps constraints to positivity bounds. By now positivity bounds have been derived in a variety of contexts, including without and with gravity, see, e.g., \cite{deRham:2017avq,deRham:2017zjm,Tolley:2020gtv,Alberte:2020bdz,Davighi:2021osh,Bellazzini:2021oaj,Chala:2021wpj} and with \cite{deRham:2017xox,deRham:2018qqo,Alberte:2019xfh,Melville:2019wyy,Alberte:2020jsk,deRham:2021fpu,Alberte:2021dnj,Herrero-Valea:2022lfd,Hamada:2023cyt} respectively, and \cite{deRham:2022hpx} for reviews.
 The impact of broken Lorentz invariance has been considered in \cite{Grall:2021xxm,Melville:2022ykg,Creminelli:2022onn} and
in a quantum gravitational context, positivity bounds have been considered in string theory \cite{Guerrieri:2021ivu,Guerrieri:2022sod,Berman:2023jys}.
\\
In order to make use of results on photon self-interactions in asymptotically safe gravity \cite{Christiansen:2017gtg,Eichhorn:2021qet,MStoappear}, we focus on positivity bounds for the photon \cite{Adams:2006sv,Henriksson:2021ymi,Haring:2022sdp,CarrilloGonzalez:2023cbf}. In the presence of gravity, the massless pole in the spin-2-propagator makes the derivation of positivity bounds more involved \cite{Alberte:2020jsk} and, as discussed, e.g., in \cite{Henriksson:2022oeu}, can lead to Planck-scale suppressed violations of the positivity bounds. Here, we nevertheless focus on the standard positivity bounds, because these can also be derived by imposing causality on the propagation of photons at low energy, i.e., without making use of scattering amplitudes \cite{CarrilloGonzalez:2023cbf}.
\\
In theories with massless vector bosons, the Lagrangian of the low-energy EFT can be written in terms of higher powers of the field strength,
	\bea
	\mathcal{L}_{\rm photons}&=&-\frac{1}{4}F_{\mu\nu}F^{\mu\nu} \\
	&{}&+ \frac{c_1}{k^4} (F_{\mu\nu}F^{\mu\nu})^2 + \frac{c_2}{k^4}F_{\mu\nu}F^{\nu \kappa}F_{\kappa\lambda}F^{\lambda\mu}  + \dots.\nonumber
	\label{eq:Wilsoncoeffs}
	\eea
Herein $k$ is an energy scale required because the higher-order terms are dimension-eight 
operators. The interactions associated to $c_{1,2}$  
form a basis for the dimension-eight 
operators at fourth 
order in field strengths 
\cite{Dittrich:2000zu, Knorr:2017kye, Eichhorn:2021qet}. Positivity bounds for the four-photon interactions
have been derived in \cite{Adams:2006sv,Henriksson:2021ymi,Haring:2022sdp,CarrilloGonzalez:2023cbf} and 
read
\be
4c_1> -3c_2,\quad \quad \frac{4c_1+3c_2}{|4c_1+c_2|}>1.\label{eq:poscons}
\ee
 The same bounds
can be derived using IR causality \cite{CarrilloGonzalez:2022fwg}, i.e., sub-luminal propagation of photons, 
in the low-energy theory, 
without reference to high-energy scattering \cite{CarrilloGonzalez:2023cbf}. 

\emph{Four-photon 
 Wilson coefficients
from asymptotically safe gravity:}
 To test, whether positivity bounds hold in asymptotically safe quantum gravity, we use functional RG techniques \cite{Wetterich:1992yh, Morris:1993qb, Ellwanger:1993mw}, see \cite{Dupuis:2020fhh} for a review.
In functional RG setups, the scale $k$ acts as an IR cutoff in the path integral, such that all quantum fluctuations at higher momentum scales have been integrated out. 
This allows a direct interpretation of $c_i(k \rightarrow 0)$ as the Wilson coefficients of a low-energy theory in which all degrees of freedom have been integrated out.  Below, we explain how to obtain predictions for these Wilson coefficients from an asymptotically safe theory of gravity and photons.

At an asymptotically safe fixed point, the higher-order interactions in Eq.~\eqref{eq:Wilsoncoeffs} are already present in the UV \cite{Christiansen:2017gtg,Eichhorn:2019yzm,Eichhorn:2021qet}. 
This is a main distinction to, e.g., asymptotically free  theories, in which such higher-order interactions are only generated by the RG flow towards the IR, but can be set to zero in the UV. 
 In asymptotically safe gravity, the higher-order interactions are generated by
 gravitational fluctuations, which we parameterize by the Einstein-Hilbert Lagrangian
\be
\label{eq:EHLagr}
\mathcal{L}_{\rm gravity} = - \frac{k^2}{16 \pi\,G} R + \frac{k^4\Lambda}{8 \pi G}.
\ee
In this parameterization, $G$ and $\Lambda$ are dimensionless and thus given by numbers in the UV.
 
 To calculate the UV values of the four-photon couplings, we make use of beta functions derived in \cite{Christiansen:2017gtg,Eichhorn:2019yzm,Eichhorn:2021qet,MStoappear}. 
The UV (fixed point) values are the zeros of the two beta functions 
\bea
\label{eq:betaschem}
\beta_{c_1}&=& 4c_1 -\frac{53}{18 \pi}G\, c_1 - \frac{23}{36 \pi} c_2 \,G + \frac{5}{2}G^2 +\mathcal{O}(c_{1,2}^2),\\
\beta_{c_2}&=& 4c_2 + \frac{19}{18 \pi}G\, c_2 + \frac{40}{9 \pi} c_1\, G- 10 G^2+\mathcal{O}(c_{1,2}^2),
\eea
where we set the UV value of the dimensionless cosmological constant to zero, $\Lambda_{\ast}=0$ (for our numerical results, we use the fixed-point value $\Lambda_{\ast}=0.05$).
At larger values of $c_{1,2}$ additional terms $\mathcal{O}(c_{1,2}^2)$ are present; they are not important for our qualitative discussion, but included in our numerical results.
The terms $\sim G^2$ imply that $\beta_{c_{1,2}}=0$ can only be achieved at $c_{1,2\, \ast} \neq 0$, i.e., at nonvanishing fixed-point values. 
To determine the fixed-point values,
we supplement the beta functions for the couplings $c_{1,2}$ by beta functions for the gravitational couplings 
calculated in \cite{Eichhorn:2016vvy}, together with the contribution from photons, as in \cite{Dona:2013qba}. 
We find the Reuter fixed point at 
\begin{equation}
\label{eq:ASFP}
G_{\ast}= 1.12\,,\quad  \Lambda_{\ast}= 0.05\,,\quad c_{1\, \ast} = -0.69\,,\quad  c_{2\, \ast}=2.62.
\end{equation}
 The two couplings $c_1$ and $c_2$ correspond to irrelevant perturbations of the Reuter fixed point, which can be determined from the critical exponents that characterize the fixed point. For irrelevant perturbations there is no free parameter that enters the values of the couplings, even once the RG flow moves away from the fixed point. 
 Because $G$ and $\Lambda$ are relevant perturbations of the fixed point, their low-energy values can be adjusted to match observational constraints \cite{Reuter:2001ag,Gubitosi:2018gsl}.  The trajectories $c_{1,2}(k)$ depend on the values of the relevant couplings, but there are no free parameters left after $G(k\ll M_{\rm Planck})$ and $\Lambda(k\ll M_{\rm Planck})$ are adjusted to their observed values.  
 We therefore obtain unique functions $c_{1,2}(k)$ and can thus determine whether or not the inequalities \eqref{eq:poscons} hold in the limit $k \rightarrow 0$.
 
For $k > M_{\rm Planck}$, it holds that $c_{1,2}(k) = c_{1,2\, \ast}$. Once $k$ drops below the Planck scale, $k < M_{\rm Planck}$, the Newton coupling scales as $G \sim  k^2$. This is necessary for gravity to behave classically in this regime, because the dimensionful Newton constant is then given by $G_N (k^2) = G\cdot k^2 =\rm const$. This fast decrease of $G$ implies that quantum fluctuations of gravity decouple dynamically, as one would expect in this regime.  At $k < M_{\rm Planck}$, the flow of four-photon couplings is driven mainly by quantum fluctuations of the photon field. 
In practise, in this regime, the RG flow of $c_{1,2}$ is dominated by dimensional terms, because $\beta_{c_{1,2}} = 4 c_{1,2} + \mathcal{O}(c_{1,2}^2)$  and $c_{1,2}$ scale towards zero.\footnote{In our setup, the scaling $G \sim k^2$ means that a tiny correction to dimensional scaling of $c_{1,2}$ is present in this regime; in settings with charged matter, this would be completely negligible compared to the effect of matter loops.} 

To interpret the resulting RG trajectories $c_{1,2}(k)$ in the context of positivity bounds, we must connect the calculation of $c_{1,2}(k)$ in a Euclidean regime to the Lorentzian positivity bounds.
We thus perform a Wick-rotation, under which the higher-order couplings do not change signs. While a Wick-rotation is in general not available in a fully non-perturbative quantum-gravity setting, see, e.g., \cite{Baldazzi:2018mtl}, there is mounting evidence for a near-perturbative nature of asymptotically safe quantum gravity \cite{Niedermaier:2009zz,Falls:2013bv, Falls:2014tra, Denz:2016qks, Falls:2017lst, Eichhorn:2018akn, Falls:2018ylp, Eichhorn:2018ydy, Eichhorn:2018nda, Eichhorn:2020sbo,Kluth:2022vnq,Becker:2024tuw}. In such a near-perturbative setting, the theory may be dominated by small fluctuations of the metric about a Minkowski background, for which an analytical continuation is available, given a suitable analyticity structure of the propagator. We thus make the assumption that a Wick-rotation of the effective action for photons captures the behavior of an asymptotically safe gravity-photon system in Lorentzian signature.

Our results are obtained in truncations of the functional RG and are thus subject to systematic uncertainties. Therefore, it is crucial to test the robustness of our results, which we do in three different ways.
First, we include six-photon operators computed in \cite{MStoappear}, which impact the RG running of the four-photon operators. They have a subleading impact on our results at the quantitative level and do not impact the qualitative results. This is in line with the expected near-perturbative nature of the fixed point, at which higher-order operators contribute at a subleading level.
Second, we vary an unphysical parameter in our calculation, namely a gauge parameter in the gravitational sector. Because it is an unphysical parameter, physical results -- such as whether causality can be violated -- would not depend on it, if we could calculate without introducing approximations through our truncation. The dependence on the gauge parameter is thus another measure for the robustness of our results, see, e.g., for other studies of gauge-parameter dependence \cite{Gies:2015tca,Knorr:2017fus}. Third, we use the principle of minimum sensitivity for the critical exponents as a function of the gauge parameter to find a preferred value for the gauge parameter. We use this value to present the results in this letter and discuss other values and provide further details in the supplementary material.

\emph{Positivity bounds in asymptotically safe gravity-photon systems:}
\begin{figure}[!t]
\includegraphics[width=\linewidth]{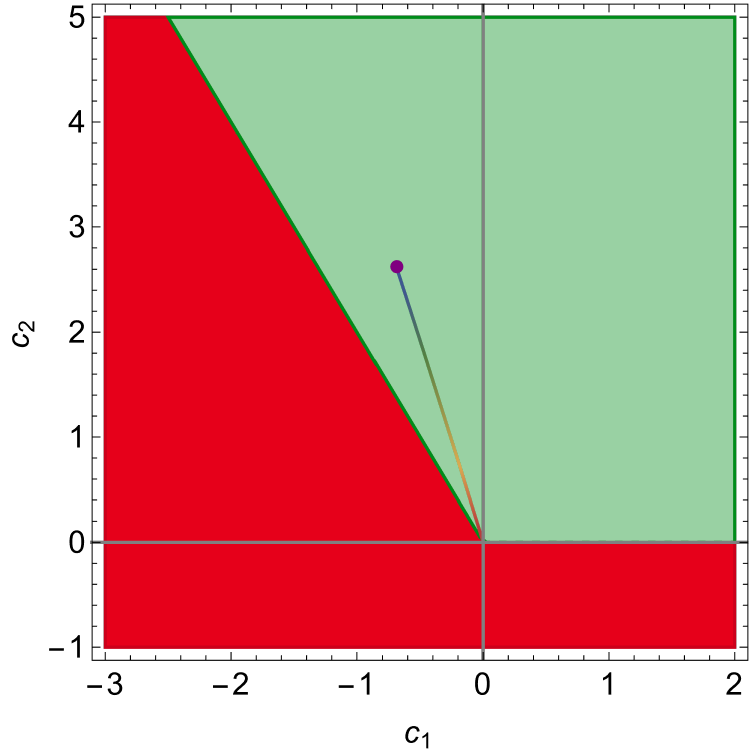}
\caption{\label{fig:c1c2} We show the RG trajectory in the $c_{1,2}$ plane; the scale dependence of the other couplings is not included in the plot, but enters in the calculation of $c_{1,2}(k)$. The purple dot indicates the Reuter fixed point of asymptotically safe quantum gravity Eq.~\eqref{eq:ASFP}, and the colored line emanating from it the RG trajectory towards $k\to0$. The green shaded region indicates where the positivity bounds \eqref{eq:poscons} are satisfied. We observe that for all finite $k$ the scale-dependent couplings $c_{1,2}(k)$ satisfy the positivity bounds.}
\end{figure}

We find that the inequalities Eq.~\eqref{eq:poscons} are satisfied at the asymptotically safe fixed point Eq.~\eqref{eq:ASFP}.
However, the decisive test of whether asymptotically safe gravity is compatible with the positivity bounds -- and thus the underlying conditions of unitarity, (micro)causality, locality and Lorentz symmetry --  is performed by starting the RG flow at the fixed point and integrating towards low $k$. We find that the positivity bounds are also satisfied everywhere along the trajectories $c_{1,2}(k)$, cf.~Fig.~\ref{fig:c1c2}. This is a non-trivial indication for the physical viability of asymptotically safe gravity.\\
In the supplemental material, we analyze the robustness of this result and find remarkable stability under extensions of the truncation and variations of the gauge parameter.
 In fact, positivity bounds hold in all settings we analyze, except for 
 a  small interval of values of the gauge parameter. We argue that this interval is disfavored, see the discussion in the supplemental material. Thus, while our study cannot fully exclude positivity violations, they  likely only arise as a consequence of an insufficient truncation for the interval of gauge parameter in question.

Accordingly, we suggest that infrared causality is preserved in the EFT for photons (see  \cite{CarrilloGonzalez:2023cbf}) whose UV completion is an asymptotically safe gravity-photon theory. In addition, the following statement can be made about unitarity, microcausality, locality and Lorentz invariance: because these conditions together imply the positivity bounds, a  large violation of positivity bounds implies that at least one of the conditions is violated.\footnote{As discussed in \cite{Henriksson:2022oeu,CarrilloGonzalez:2023cbf}, Planck-scale suppressed violations of positivity bounds can occur in the presence of gravity.} The converse is not true; if positivity bounds are satisfied, one cannot conclude that all conditions hold, as they might, e.g., be violated in positivity bounds
 for higher-order Wilson coefficients
beyond those we considered. We can thus make the statement that \emph{we do {\bf not} find convincing indications for the violation of unitarity, locality, microcausality or Lorentz-symmetry in asymptotically safe gravity}.\\

\begin{figure}[!t]
\includegraphics[width=\linewidth]{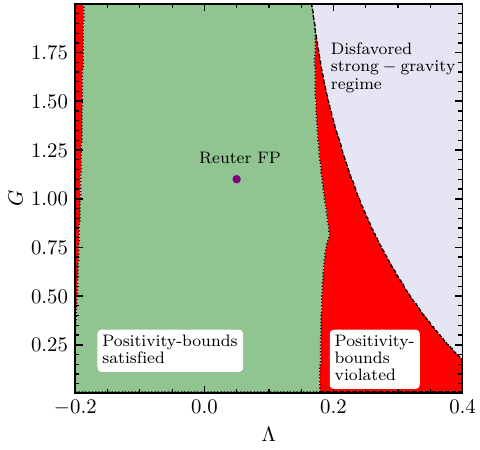}
\caption{\label{fig:GLambda} Region in the $G-\Lambda$ plane, in which the positivity bounds on $c_{1,2}$ are satisfied at $ k\to0$. The gravitational couplings are treated as input parameters, and their scale-dependence is approximated by simple Heaviside functions, which is a good approximation of the actual scale dependence. The green region indicates where the positivity bounds are satisfied, while the red region indicates where they are violated. The gray region indicates a region where new relevant directions are present in the matter sector. The purple dot indicates the Reuter fixed point obtained within a specific setup.}
\end{figure}

Next, we explore whether positivity bounds automatically hold in asymptotically safe gravity or whether they constrain the theory in a non-trivial way. To that end, we treat the fixed-point values of the gravitational couplings $G$ and $\Lambda$ as free parameters, instead of determining them through the fixed-point condition. 
This analysis can also be interpreted as a further test of robustness of our results: because the fixed-point values of $G$ and $\Lambda$ are subject to uncertainties induced by the truncation, it is crucial to know whether slight variations in the fixed-point values could lead to qualitatively different conclusions.
In the plane spanned by $G$ and $\Lambda$, we determine the fixed-point values of $c_{1,2}$ as functions of $G$ and $\Lambda$. We restrict the range of $\Lambda$ to those values that can be continuously connected to $\Lambda=0$ at low scales, see the supplementary material for details. As a first result, we find that there are regions in the $G-\Lambda$-plane for which the positivity bounds are violated for the fixed-point values $c_{1,2\, \ast}(G,\Lambda)$. This does not automatically imply positivity violation in the effective field theory, because the positivity bounds relate to the Wilson coefficients, i.e., the couplings at low $k$, and not their fixed-point values. To test positivity violation, we numerically solve the beta functions to obtain $c_{1,2}(k)$ and calculate $c_{1,2}(k \rightarrow 0)$. For this analysis we approximate the scale-dependent gravitational couplings $G(k)$ and $\Lambda(k)$ with Heaviside functions, i.e., they assume the input values $(G,\Lambda)$ above the Planck-scale, and are zero below the Planck scale. This approximation is justified, since $G\sim k^2$ and $\Lambda(k\to0)\approx0$, and hence very quickly approach zero below the Planck scale. We also check explicitly that for the case, where $G(k)$ and $\Lambda(k)$ are described by fixed-point trajectories, this approximation leads to the same conclusion regarding positivity bounds. \\
The result of this analysis is shown in Fig.~\ref{fig:GLambda}. There is a region in the $G-\Lambda$-plane, in which the positivity bounds are satisfied. The Reuter fixed point is located  inside this region and well away from its boundaries.
 There is also a region, in which the positivity bounds are violated, as well as a region in which some truncations do not show a fixed point and other truncations only show fixed points at which higher-order couplings become relevant. This region is referred to as ``disfavored strong-gravity regime'', because the non-existence of fixed points or presence of extra relevant directions is tied to gravity fluctuations becoming nonperturbatively strong, see \cite{Christiansen:2017gtg,Eichhorn:2019yzm,Eichhorn:2021qet, MStoappear, deBrito:2023myf} for details. 

From Fig.~\ref{fig:GLambda}, we can conclude that (i) it is nontrivial that the IR theory resulting from the Reuter fixed point satisfies positivity bounds, because this is not the case for all values of $G$ and $\Lambda$;
(ii) that our result on positivity bounds is robust, because the Reuter fixed point is not located close to the boundary of the region which satisfies positivity bounds; (iii) that positivity bounds can be used to place nontrivial constraints on the gravitational parameters. The third is particularly important, because without considering positivity bounds, there would not be any way (within our truncation) to see that  the assumptions going into derivations of positivity bounds may be violated for these parameter values. Only once positivity bounds are included, can we understand that not all parameter values $G, \Lambda$ outside the disfavored strong-gravity regime are actually physically viable, because some of them lead to violations of causality in the propagation of photons \cite{CarrilloGonzalez:2023cbf}.

\emph{Future perspectives:} In this work, we have pioneered a new way to detect violations of unitarity, locality, (micro-) causality or Lorentz symmetry in asymptotically safe gravity. Because in our setup positivity bounds for the lowest-order four-photon couplings in an asymptotically safe theory hold except in a small (and disfavored, see supplemental material) interval of values of a gravitational gauge parameter, we conclude that we do not find convincing evidence   for the violation of these properties. 

Our work shows that positivity bounds can be tested meaningfully in asymptotic safety, even using existing results for higher-order interactions, from which Wilson coefficients can be extracted. Several extensions of our study are technically possible and physically informative. First, the inclusion of couplings that contribute to the 2-2- photon amplitude at higher level in momenta provides further tests for the robustness of our results, but also leads to additional positivity bounds \cite{Henriksson:2021ymi,Haring:2022sdp,CarrilloGonzalez:2023cbf}. Second,  the (truncated) 2-2-photon scattering amplitude can be calculated from the Wilson coefficients, complementing the first steps towards scattering amplitudes undertaken in \cite{Draper:2020knh, Knorr:2022lzn}.  Third, positivity bounds for other matter fields, e.g., scalars \cite{deRham:2017avq}, for which higher-order terms are already rather well-explored \cite{Eichhorn:2012va, deBrito:2021pyi, Laporte:2021kyp, Knorr:2022ilz, deBrito:2023myf}, can also be explored.  
Fourth, the constraints on the gravitational parameter space can be compared with other phenomenological constraints, such as, e.g., viability of the Standard Model of particle physics \cite{Eichhorn:2016esv,Eichhorn:2017eht,Eichhorn:2017ylw,Eichhorn:2018whv}. 
Fifth, instead of working in a Euclidean regime, beta functions could be calculated either in the presence of a (Euclidean) foliation which ensures that a continuation to Lorentzian signature is more readily available, as in \cite{Manrique:2011jc, Rechenberger:2012dt, Biemans:2016rvp, Biemans:2017zca, Houthoff:2017oam, Knorr:2018fdu, Eichhorn:2019ybe, Saueressig:2023tfy, Korver:2024sam}, or even directly in Lorentzian signature \cite{Fehre:2021eob}, see also \cite{Banerjee:2022xvi, DAngelo:2022vsh, DAngelo:2023tis, DAngelo:2023wje}.
By carrying out this program, we can collect evidence as to whether asymptotically safe gravity constitutes a physically viable quantum theory of gravity. Our present work paves the way to achieve this.\\

\emph{Acknowledgements:} We thank A.~Platania, B.~Knorr, C.~de Rham and A.~Tolley for discussions.
AE is supported by a grant (no 29405) by VILLUM Fonden. AE also gratefully acknowledges support by the Deutsche Forschungsgemeinschaft (DFG) under Grant No 406116891 within the Research Training Group RTG 2522/1 during an early stage of this work. This work is also supported by the Perimeter Institute for Theoretical Physics. Research at Perimeter Institute is supported in part by the Government of Canada through the Department of Innovation, Science and Economic Development and by the Province of Ontario through the Ministry of Colleges and Universities.

\newpage

\section{Supplementary material}

Here, we investigate the robustness of our results. We use the principle of minimum sensitivity to identify a regime of the gauge parameter in which results are stable under extensions of the truncation and only very weakly dependent on the gauge parameter. Working in this regime, we then show how the removal of $F^6$ operators from our study has only a subleading, quantitative impact on our results, but does not alter the qualitative conclusions. Furthermore, we investigate the gauge-dependence of the positivity bounds in our system, showing that indeed the violation or non-violation of positivity bounds is stable under small changes of the gauge-parameter in the regime singled out by the principle of minimal sensitivity.

\subsection{Technical setup and gauge choice}
For completeness, let us first provide some details on the technical setup. We use the functional renormalization group (FRG) \cite{Wetterich:1992yh, Morris:1993qb, Ellwanger:1993mw,Reuter:1996cp} to extract the scale-dependence of the couplings in the photon-gravity system. The key object of the FRG is the scale-dependent effective action $\Gamma_k$, which includes all quantum fluctuations above the IR-cutoff scale $k$. The FRG furthermore provides a functional differential equation, 
	\begin{equation}
		\label{eq:floweq}
		k\partial_k \Gamma_k=\frac{1}{2}\mathrm{Tr}\left[k\partial_k R_k \left(\Gamma_k^{(2)}+R_k\right)^{-1}\right]\,,
	\end{equation}
	where $\Gamma_k^{(2)}$ is the second derivative of $\Gamma_k$ with respect to the fields, and where $R_k$ is the IR regulator. In the path integral, it acts akin a momentum-dependent mass, and hence implements the Wilsonian idea of integrating out quantum fluctuations according to their momentum scale. From the flow equation Eq.~\eqref{eq:floweq}, we can extract the scale dependence of all couplings of the system, by projecting onto suitable field monomials or functionals. While $\Gamma_k$ in principle contains all possible operators that are compatible with the symmetries of the system, in practical computations, it is typically truncated to a finite set of operators. Systematic extensions of this set of operators are then necessary to i) estimate the systematic uncertainties in a given truncation, and ii) to obtain qualitatively reliable results of physical observables.
	For detailed reviews on the FRG, and its applications in gravity see, e.g., \cite{Dupuis:2020fhh,Reichert:2020mja,Saueressig:2023irs,Pawlowski:2023gym}.\\
	In the present work we approximate the gravitational dynamics by the Einstein-Hilbert action, see Eq.~\eqref{eq:EHLagr}, and the dynamics of the photon sector with
	\begin{align}
		\label{eq:gammakone}
		\Gamma_k^{\mathrm{photons}} =& \frac{1}{4} \Int 	\mathcal{F}_2\,   \nonumber\\
		& + \Int \left( \frac{c_1}{k^4}\, \left(	\mathcal{F}_2\right)^2 +\frac{c_2}{k^4}	\mathcal{F}_4\right) \nonumber\\
		& + \Int \left( \frac{d_1}{k^8}\, \left(	\mathcal{F}_2\right)^3 +\frac{d_2}{k^4} \mathcal{F}_2\,\mathcal{F}_4\right)\,, 
	\end{align}
	where we define
	\bea
		\mathcal{F}_2&=&g^{\mu\nu}g^{\kappa\lambda}F_{\mu\kappa}F_{\nu\lambda}\,,\\
				\mathcal{F}_4&=&g^{\mu\nu}g^{\kappa\lambda}g^{\rho\sigma}g^{\zeta \tau}F_{\mu\kappa}F_{\lambda \rho}F_{\sigma \zeta}F_{\tau \nu},
	\eea
	which is the leading-order photon EFT-action Eq.~\eqref{eq:Wilsoncoeffs}, supplemented with two six-photon operators, corresponding to $d_{1,2}$.
	The interactions associated to $c_{1,2}$  ($d_{1,2}$) form a basis for the dimension-eight (twelve) operators at fourth (sixth) order in field strengths without additional derivatives, which can be shown using the relations in \cite{Dittrich:2000zu, Knorr:2017kye, Eichhorn:2021qet}. 
	From a phenomenological perspective, we are mainly interested in the scale dependence of the four-photon interactions $c_{1,2}(k)$ \cite{Christiansen:2017gtg,Eichhorn:2019yzm,Eichhorn:2021qet}, and we will use the six-photon interactions, computed in \cite{MStoappear}, to assess the robustness of our results. To this end, we will refer to a truncation where we set $d_{1,2}=0$ as '$F^4$-truncation', while we refer to the system including the interactions associated to $d_{1,2}$ as '$F^6$-truncation'. 
	
	To extract the scale dependence of couplings, we expand the action around an auxiliary background, using a linear parameterization of metric fluctuations, i.e.,
		\begin{equation}
			g_{\mu\nu}=\bar{g}_{\mu\nu}+\sqrt{16\,\pi k^{-2} G(k)} h_{\mu\nu}\,.
		\end{equation}
	In order to compute the impact of metric and photon fluctuations, we need to fix the gauge in both sectors. Hence, we introduce the gauge-fixing actions
	\begin{align}
		S_{\mathrm{gf},\, h} &= \frac{1}{32\,\pi\, \alpha_h\, G(k)\, k^{-2}} \Intb \mathcal{F}^{\mu} \bar{g}_{\mu\nu}\mathcal{F}^{\nu}\,,
	\end{align}
	with
	\begin{equation}
		\mathcal{F}^{\mu} = \left(\delta^{\mu\kappa}\bar{D}^{\lambda} - \frac{1+ \beta_h}{4} \delta^{\kappa\lambda}\bar{D}^{\mu}\right) h_{\kappa\lambda}\,,
	\end{equation}
	and 
	\begin{equation}
		S_{\mathrm{gf},\,A} = \frac{1}{2\alpha_A} \Intb\left(\bar{D}^{\nu}A_{\nu}\right)\left(\bar{D}^{\mu} A_{\mu}\right)\,,
	\end{equation}
	where we have introduced a total of three gauge parameters $\alpha_h$, $\alpha_A$, and $\beta_h$. The first two control how strictly the gauge condition is imposed, while $\beta_h$ determines which of the scalar modes in the metric-sector is physical. We choose the Landau gauge in both sectors, i.e.,
	\begin{equation}
		\alpha_h\to0\,,\quad\text{and}\quad \alpha_A\to0\,,
	\end{equation}
	which has been shown to be a preferred choice \cite{Litim:2002ce, Knorr:2017fus}.
	We then extract the scale dependence of the photon interactions $c_{1,2}$, and $d_{1,2}$ as a function of $\beta_h$, and take the scale dependence of $G$ and $\Lambda$ as a function of $\beta_h$ from \cite{Eichhorn:2016vvy}, supplemented with the photon contribution from \cite{Dona:2013qba}. We note that in the presence of gravity, couplings such as $c_{1,2}$ and $d_{1,2}$ are automatically induced in the UV, and cannot be consistently set to zero, see also Eq.~\eqref{eq:betaschem}. More generally speaking, there is strong evidence that at an asymptotically safe fixed point for gravity and matter, all couplings that respect the symmetries of the system, including matter self-interactions \cite{ Eichhorn:2011pc, Eichhorn:2012va, Meibohm:2016mkp,  Eichhorn:2016esv, Eichhorn:2017eht, Christiansen:2017gtg, Eichhorn:2019yzm, deBrito:2020dta,   Laporte:2021kyp,  deBrito:2021pyi, Eichhorn:2021qet, Knorr:2022ilz, deBrito:2023myf, deBrito:2023kow} and non-minimal interactions \cite{ Eichhorn:2017sok, Eichhorn:2017eht, Eichhorn:2018nda, Laporte:2021kyp, Knorr:2022ilz} are non-vanishing. In turn, those operators that break a symmetry of the system can consistently be set to zero, see, e.g., \cite{Narain:2009fy, Percacci:2015wwa, Eichhorn:2017als,  Eichhorn:2016vvy, Eichhorn:2017eht, Knorr:2018fdu, Eichhorn:2019ybe}. 
	
\begin{figure}[!t]
		\includegraphics[width=\linewidth]{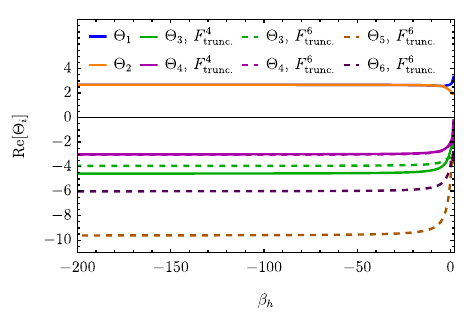}
		\caption{\label{fig:gaugedep} 
		Critical exponents at the fixed point of the gravity-photon system for different values of the gauge parameter $\beta_h$. $\Theta_{1,2}$ correspond to the critical exponents of the gravitational couplings $G$ and $\Lambda$, and are hence independent of the truncation in the matter sector. The solid (dashed) lines at negative $\Theta_i$ indicate the critical exponents of the matter couplings in the $F^4$- and the $F^6$-truncation. Positive critical exponents indicate relevant directions and negative critical exponents irrelevant ones.}
	\end{figure}

	While the flow equation Eq.~\eqref{eq:floweq} is in principle exact, within truncation also physical quantities, such as critical exponents, depend on the  gauge parameters of the gravitational sector. Critical exponents define the universality class of a fixed point and are defined by
	\begin{equation}
		\Theta_i=-\mathrm{eig}\left[\frac{\partial \beta_{g_i}}{\partial g_{j}}\right]\bigg|_{g_l=g_{l,\,*}}\,,
	\end{equation}
	where $g_{i}$ are all couplings of the system.
	In order to minimize the residual gauge dependence of our truncated system, we apply the principle of minimal sensitivity (PMS) to the critical exponents of the system in order to choose an appropriate value for $\beta_h$. The PMS has successfully been employed in the context of the FRG to minimize residual regulator dependence in \cite{Balog:2019rrg}, and in quantum gravity in \cite{Baldazzi:2023pep}.
	In \autoref{fig:gaugedep} we show the set of critical exponents in the $F^4$- and the $F^6$-truncation as a function of the gauge parameter $\beta_h$. Here, $\Theta_{1,2}$ correspond to the critical exponents of the gravitational couplings. In our system, they are not impacted by the truncation in the matter sector. We see that both the gravitational and the photon critical exponents vary strongly around $\beta_h=0$, and become constant towards $\beta_h\to-\infty$. This is the case both in the $F^4$- (solid lines), and the $F^6$- truncation (dashed lines). This indicates that negative values of $\beta_h$ are preferred by a PMS on the gauge parameter. Hence, we choose 
	\begin{equation}
		\beta_h\to-\infty\,,
		\end{equation}
	 for the analysis presented in this paper.  In this gauge, the beta-functions feature poles at $\Lambda=-\tfrac{1}{4}$ and $\Lambda=\tfrac{1}{2}$, which correspond to the poles of the scalar and the transverse-traceless mode of metric fluctuations. This pole structure limits the range of admissible fixed-point values to $\Lambda\in(-\tfrac{1}{4},\tfrac{1}{2})$, which are the only values that can result in a nearly vanishing cosmological constant at low scales, in the present approximation. Note that a more careful study of gravitational dynamics would treat the avatar of the cosmological constant appearing in the propagator differently from the coupling appearing as the momentum-independent part of pure-gravity vertices, see, e.g., \cite{Christiansen:2014raa, Christiansen:2015rva, Pawlowski:2020qer}. 
	 
\begin{figure}[!t]
	\includegraphics[width=\linewidth]{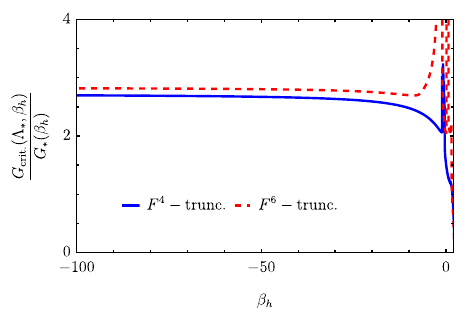}\\
	\includegraphics[width=\linewidth]{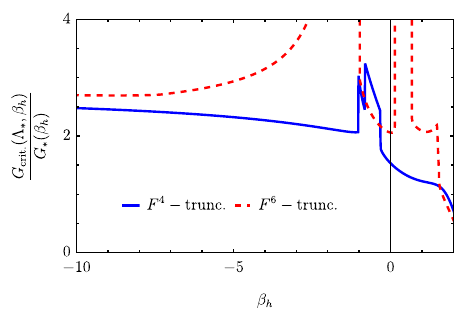}
		\caption{\label{fig:gfpgcritf4f6}  We show the ratio $G_{\rm crit}(\Lambda_{\ast})/G_{\ast}$. At $G_{\rm crit}$, the fixed point for the four-photon couplings is lost to a fixed point collision in the $F^4$ truncation (blue continuous line) and acquires a relevant direction in the $F^6$ truncation (red dashed line). Because the difference between a fixed-point collision and a real fixed point with an extra relevant direction is major, we consider values $G> G_{\rm crit}$ unreliable in the present truncation. Thus, if the fixed-point value $G_{\ast}$ gets too close to $G_{\rm crit}$, we no longer fully trust the results obtained in these comparatively small truncations. The jumps are caused by jumps in $G_{\mathrm{crit.}}$, or the overall disappearence of disfavored regions for specific gauge parameters $\beta_h$.}
\end{figure}

Anticipating a study of other values of the gauge parameter below, we explore whether our truncation appears equally robust for all values of the gauge parameter.\footnote{ Let us highlight that we explore $\beta_h$ at a fixed value $\alpha=0$; a study of other values of $\alpha$, e.g., $\alpha=1$ as used in \cite{ABtoAppear} would also be highly informative and is postponed to future work.} We do so by comparing the fixed-point value of the Reuter fixed point, $G_{\ast}$, to the value $G_{\rm crit}$, at which the disfavored strong-gravity regime begins. In this regime, our lowest-order truncation does not feature a fixed point for the four-photon couplings, whereas a larger truncation features a fixed point, but with additional relevant couplings. Such drastic changes under extensions of the truncation indicate that more extended truncations are needed to achieve robust results which are (apparently) converged. Furthermore, in scalar-tensor theories, it has been explicitly explored that truncations close to such a disfavored regime are not converged yet in small truncations, see \cite{deBrito:2023myf}.
By comparing $G_{\rm crit}/G_{\ast}$ in the $F^4$ and $F^6$ truncation, we see that the system appears stable under extensions of the truncation at large negative $\beta_h$, but not stable for $\beta_h \gtrsim -6$, cf.~Fig.~\ref{fig:gfpgcritf4f6}, because the ratio $G_{\rm crit}/G_{\ast}$ changes strongly between the two truncations in this regime. 
	 
\subsection{Stability under an extension of the truncation}
	\begin{figure}[!t]
	\includegraphics[width=\linewidth]{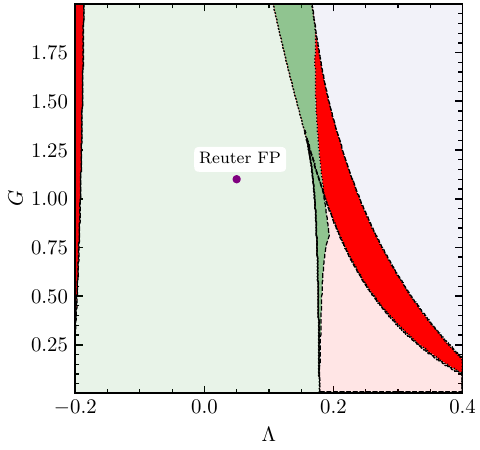}
	\caption{\label{fig:F4vsF6IR} 
		Comparison of regions where positivity bounds are satisfied or violated in the $F^4$- and the $F^6$-truncations. If a region has changed its status from the $F^4$-to the $F^6$-truncation, it is indicated in dark green (red) if the positivity bound is satisfied (violated) in the $F^6$- truncation. Light colors indicate regions where both truncations lead to the same result, and light green (red) indicate where the positivity bounds for $c_{1,2}$ are satisfied (violated) in both truncations.}
\end{figure}
To assess the robustness of our analysis, we compare the regions where positivity bounds for $c_{1,2}$ are satisfied and violated in the $F^4$- and $F^6$-truncation, and within the Heaviside approximation for the scale-dependence of $G$ and $\Lambda$. The $F^4$-truncation is the lowest order at which positivity bounds in the photon sector of asymptotically safe quantum gravity can be tested. In computations based on the FRG, the scale dependence of $n$-point functions generically depends on $n+1$ and $n+2$-point functions, which in this case includes six-photon interactions. These additional interactions  impact $c_{1,2}$ in two ways: first, the asymptotically safe fixed point changes, and second, the flow towards $k\to0$ is modified, resulting in a modified trajectory of the RG-flow, ultimately resulting in changed Wilson coefficients at $k\to0$.

Hence, if the regions where positivity bounds are satisfied change significantly under the inclusion of $F^6$ operators, this indicates that our truncation is not yet converged and any statements on positivity bounds being satisfied or violated come with a large systematic uncertainty. Conversely, if the status of positivity bounds only changes in small regions in the $G$-$\Lambda$ plane when increasing the truncation, this indicates some amount of apparent convergence, and reliability of our truncation.

In Fig.~\ref{fig:F4vsF6IR}, we show which regions in the $G$-$\Lambda$ plane change their status regarding positivity bounds from the $F^4$-  to the $F^6$-truncation. Here, dark green (red) indicates that positivity bounds are satisfied (violated) in the $F^6$-truncation, but have not been satisfied (violated) in that region in the $F^4$-truncation. Light colors (green, red, blue), indicate regions where the status of positivity bounds has not changed from the $F^4$- to the $F^6$-truncation. We observe that a large region in the gravitational parameter space is colored lightly, meaning that the status of positivity bounds is stable under extensions of the truncation. Only narrow bands have changed their status. Among those regions, only a small band around $\Lambda\approx -0.2$ violates positivity in the $F^6$-truncation, while positivity bounds are satisfied in the $F^4$-truncation in that region. In the other regions, positivity either changes from violated to satisfied, or from disfavored strong-gravity regions to either satisfaction or violation of positivity bounds in the $F^6$-truncation. 

\subsection{Stability under changes of the gauge parameter}\label{sec:gaugeparam}
\begin{figure}[!t]
	\includegraphics[width=\linewidth]{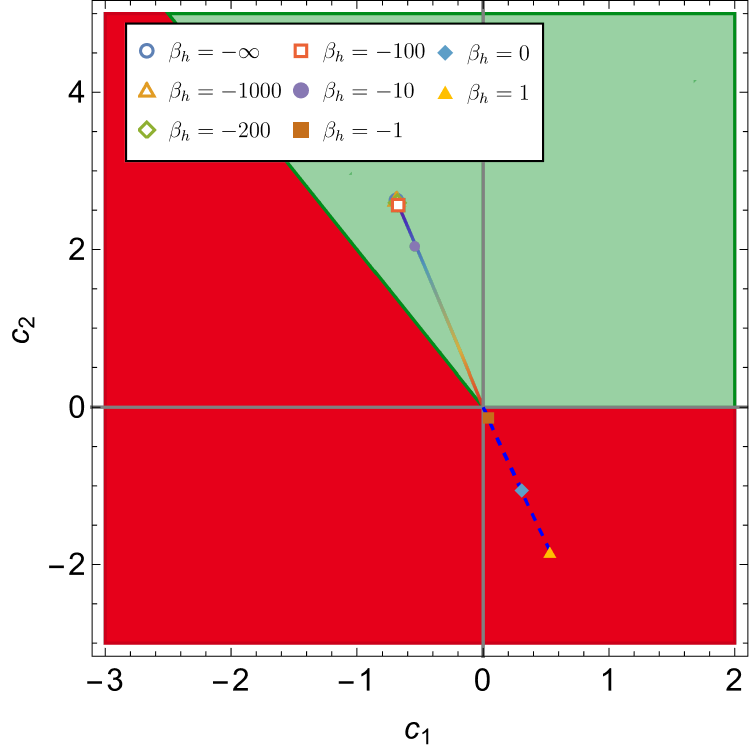}
	\caption{\label{fig:GaugePedc1c2Traj} 
		We show the RG trajectory in the $c_{1,2}$ plane; the scale dependence of the other couplings is not included in the plot, but enters in the calculation of $c_{1,2}(k)$. The markers indicates the Reuter fixed point of asymptotically safe quantum gravity Eq.~\eqref{eq:ASFP} for different choices of the gauge parameter $\beta_h$. The colored solid (blue dashed) line indicates the RG trajectory towards $k\to0$ for $\beta_h\to-\infty$ ($\beta_h=1$). The green shaded region indicates where both positivity bounds Eq.~\eqref{eq:poscons} are satisfied. The  red region indicates where one of the positivity bounds is violated. We observe that for all finite $k$, the scale-dependent couplings $c_{1,2}(k)$ either satisfy  or violate the positivity bounds, depending on the quadrant the fixed-point value lies in.}
\end{figure}
As mentioned before, in truncations of the scale dependent effective action $\Gamma_k$, also physical quantities, like critical exponents,  depend on unphysical parameters, such as the gauge parameter $\beta_h$. We have argued before that a principle of minimal sensitivity applied on the critical exponents of the Reuter fixed point singles out $\beta_h\to-\infty$ as a preferred gauge choice, and all results presented in the main part of this paper were obtained for that choice. Here we focus on another physical quantity, namely the question whether positivity bounds (at $k\to0$) are satisfied or not for theories emanating from the Reuter fixed point. 

In Fig.~\ref{fig:GaugePedc1c2Traj} we show the RG trajectories emanating from the Reuter fixed point for different choices of $\beta_h$ in the $c_{1,2}$ plane. The markers indicate the location of the Reuter fixed point, while the colored solid line (the blue dashed line) indicates the trajectories for $\beta_h\to-\infty$ ($\beta_h=1$). We see that $c_{1 *}$ and $c_{2*}$ exchange their sign at about $\beta_h\approx-1$. Furthermore, we see that for all shown values of $\beta_h$, the Reuter fixed point results in an almost constant value for $\frac{c_{1,*}}{c_{2,*}}$. Correspondingly, for $\beta_h\gtrsim-1$, both positivity-bounds Eq.~\eqref{eq:poscons} are violated along the entire RG-trajectory, while for $\beta_h\lesssim-1$ the positivity bounds are satisfied along the entire RG-trajectory.%

\begin{figure}[!t]
	\includegraphics[width=\linewidth]{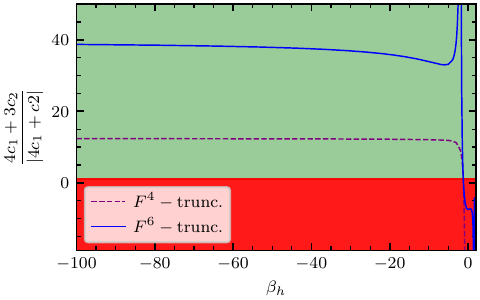}
	\caption{\label{fig:GaugePosBd} 
		We show the value of the stronger positivity condition in Eq.~\eqref{eq:poscons} at $k\to0$ for trajectories emanating from the Reuter fixed point as a function of the gauge parameter $\beta_h$ in the $F^4$- (purple dashed line) and the $F^6$-truncation (blue solid line). For the scale dependence of the gravitational couplings we used the Heaviside approximation. The green region indicates $\frac{4c_1+3c_2}{|4c_1+c_2|}>1$, i.e., satisfied positivity bounds, while the red region marks $\frac{4c_1+3c_2}{|4c_1+c_2|}<1$.
		}
\end{figure}

\begin{figure}[!t]
	\includegraphics[width=\linewidth]{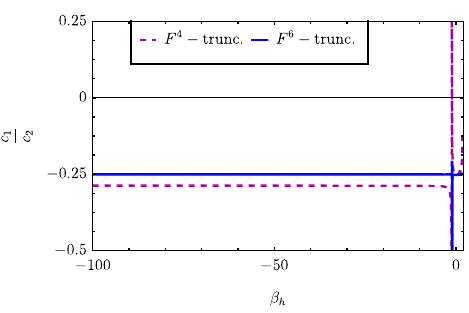}
	\caption{\label{fig:WCoeffRat} 
		We show the ratio of Wilson coefficients $c_{1,2}$ at $k\to0$ for trajectories emanating from the Reuter fixed point as a function of the gauge parameter $\beta_h$. For the scale dependence of the gravitational couplings we used the Heaviside approximation. We see that this ratio is very stable as a function of $\beta_h$, except for a divergence around $\beta_h\approx-1$, where $c_1$ and $c_2$ change sign shortly after one another. Since the sign-change does not occur at exactly the same value for $\beta_h$, the ratio diverges. We also see quantitative agreement between the $F^4$- and the $F^6$-truncation, both in the value of the ratio, as well as in the location of the sign-change.
	}
\end{figure}

In Fig.~\ref{fig:GaugePosBd} we show the value of the stronger positivity bound at $k\to0$ for trajectories emanating from the Reuter fixed point, and as a function of $\beta_h$. We employed the Heaviside approximation for the scale dependence of the gravitational couplings.   We see that indeed the combination  $\frac{4c_1+3c_2}{|4c_1+c_2|}$ is very stable for negative $\beta_h$, and only changes significantly around $\beta_h\approx-4$, where it then drops below 1 around $\beta_h\approx-1$, indicating the violation of positivity bounds. This behavior, namely the stability for large negative $\beta_h$, and rapid changes for $\beta_h\approx0$ are in line with the behavior of critical exponents, see \autoref{fig:gaugedep}. 
While this holds independently for the $F^4$- and the $F^6$- truncation, we see a large quantitative difference between both truncations.  However, as we see in \autoref{fig:WCoeffRat}, this it not caused by a large truncation-dependence of the Wilson coefficients $c_{1,2}$, but by a near-cancellation in the denominator: we find that the approximation $4c_1\approx-c_2$ is very stable as a function of $\beta_h$, up to $\beta_h\approx-1$. Furthermore, we see that this approximation becomes more accurate in the $F^6$-truncation. Since the combination $4c_1+c_2$ is precisely the denominator of the positivity bound Eq.~\eqref{eq:poscons}, the value of the positivity condition, which is shown in \autoref{fig:GaugePosBd}, increases significantly, due to the denominator being close to zero.

We conclude that there are gauges where RG trajectories emanating from the corresponding Reuter fixed points violate the positivity bounds, but that those gauge choices lie within (or very close to) the range that we have conservatively excluded above.

\bibliography{references.bib, refs.bib}
\end{document}